# Spinor quintom cosmology with intrinsic spin

## Emre DİL

Department of Physics, Sinop University, 57000, Korucuk, Sinop-TURKEY

**Abstract:** We consider a spinor quintom dark energy model with intrinsic spin, in the framework of Eintein-Cartan-Sciama-Kibble theory. After constructing the mathematical formalism of the model, we obtain the spin contributed total energy-momentum tensor giving the energy density and the pressure of the quintom model, and then we find the equation of state parameter in terms of the spinor potential. Choosing suitable potentials leads to the quintom scenario crossing between quintessence and phantom epochs, or vice versa. Analyzed three quintom scenarios provides stable expansion phases avoiding Big Rip singularities, and yielding matter dominated era through the stabilization of the spinor pressure via spin contribution. The stabilization in spinor pressure leads to neglecting it as compared to the increasing energy density, and constituting a matter dominated stable expansion epoch.

**PACS No:** 95.36.+x; 98.80.-k; 11.25.-w; 04.50.Kd

**Keywords:** Dark energy; Quintom field; Dirac Spinor; Eintein-Cartan gravity

## 1. Introduction

The astrophysical observations show that the universe is experiencing an accelerating expansion due to an unknown component of energy, named as the dark energy (DE) which is distributed all over the universe and having a negative pressure in order to drive the acceleration of the universe [1-9]. There have been proposed various DE scenarios: Cosmological constant $\Lambda$ is the oldest DE model which has a constant energy density filling the space homogeneously [10-13]. Equation of state (EoS) parameter of a cosmological fluid is $\omega = p/\rho$, where $p$ and $\rho$ are the pressure and energy density and the DE scenario formed by the cosmological constant refers to a perfect fluid with EoS $\omega_\Lambda = -1$.

Other DE scenarios can be constructed from the dynamical components, such that the quintessence, phantom, K-essence, or quintom [14-16]. Quintessence is considered as a DE scenario with Eos $\omega > -1$. Such a model can be described by using a canonical scalar field. Recent observational data presents that the EoS of DE can be in a region where $\omega < -1$. The most common scenario generalizing this regime is to use a scalar field with a negative kinetic term. This DE model is known as the phantom scenario which is also named as a ghost [17]. This model experiences the shortcoming, such that its energy state is unbounded from below and this leads to the quantum instability problem [18]. If the potential value is not bounded from above, this scenario is even instable at the classical regime known as the Big Rip Singularity [19]. Another DE scenario, K-essence is constructed by using kinetic term in the domain of a general function in the field Lagrangian. This model can realize both $\omega > -1$ and $\omega < -1$ due to the existence of a positive energy density, but cannot allow a consistent crossing of the cosmological constant boundary $-1$ [20].

The time variation of the EoS of DE has been restricted by the data obtained by Supernovae Ia (SNIa). According to the SNIa data, some attempts have come out to estimate the band power and density of the DE EoS as a function of the redshift [21]. There occurs two main models for the variation of EoS with respect to time; Model A and Model B. While the Model A is valid in low redshift, the Model B suffers in low redshift; therefore it works in high redshift values. These models implies that the evolution of the DE EoS begins from $\omega > -1$ in the past to $\omega < -1$ in the present time, namely the observational and theoretical results allows the EoS $\omega$ of DE crossing the cosmological constant boundary or phantom divide during the evolution of the universe [22-25].

Crossing of the DE EoS the cosmological constant boundary is named as the Quintom scenario and this can be constructed in some special DE models. For instance, if we consider a single perfect fluid or single scalar field DE constituent, this model does not allow the DE EoS cross the $\omega = -1$ boundary according to the no-go theorem [26-31]. To overcome no-go theorem and to realize crossing phantom divide line, some modifications can be made to the single scalar field DE models. One can construct a quintom scenario by considering two scalar fields, such as quintessence and quintom [32]. The components cannot cross the $-1$ boundary alone, but can across it combined together. Another quintom scenario is achieved by constructing a scalar field model with non-linear or higher order derivative kinetic term [33,34] or a phantom model coupled to dark matter [35]. Also the scalar field DE models non-minimally coupled to gravity satisfy the crossing cosmological constant boundary [36-37].

The aforementioned quintom models are constructed from the scalar fields providing various phantom behaviors, but the ghost field may cause some instable solutions. By considering the linearized perturbations in the effective quantum field equation at two-loop order one can obtain an acceleration phase [38-42]. On the other hand, there is another quintom model satisfying the acceleration of the universe, which is constructed from the classical homogeneous spinor field $\psi$ [43-45]. In recent years, there can be found many studies for spinor fields in cosmology [20], such that, for inflation and cyclic universe driven by spinor fields, for spinor matter in Bianchi Type I spacetime, for a DE model with spinor matter [46-51].

The consistent quintom cosmology has been proposed by using spinor matter in Friedmann-Robertson-Walker (FRW) geometry, in Einstein's general relativity framework [52]. The spinor quintom scenario allows EoS crossing $-1$ boundary without using a ghost field. When the derivative of the potential term with respect to the scalar bilinear $\bar{\psi}\psi$ becomes negative, the spinor field shows a phantom-like behavior. But the spinor quintom exhibits a quintessence-like behavior for the positive definite potential derivative [20]. In this quintom model, there exist three categories of scenario depending on the choice of the type of potentials; one scenario is that the universe evolve from a quintessence-like phase $\omega > -1$ to a phantom-like phase $\omega < -1$, another scenario is for the universe evolving from a phantom-like phase $\omega < -1$ to a quintessence-like phase $\omega > -1$, and the third scenario is that the EoS of spinor quintom DE crosses the $-1$ boundary more than one time.

In this study, we consider the spinor quintom DE, in the framework of Eintein-Cartan-Sciama-Kibble (ECSK) theory which is a generalization of the metric-affine formulation of the Einstein's general relativity with intrinsic spin [53-62]. Since the ECKS theory is the simplest theory including the intrinsic spin and avoiding the big-bang singularity [63], it is worth considering the spinor quintom in ESCK theory for investigating the acceleration phase of the universe with the phantom behavior. Therefore, we analyze the spinor quintom model with intrinsic spin in ECSK theory whether it provides the crossing cosmological constant boundary. Then if the model provides the crossing $-1$ boundary, we will find the suitable conditions on the potential for the crossing $-1$ boundary.

## 2. Algebra of spinor quintom with intrinsic spin

The most complicated example of the quantum field theories lying in curved spacetime is the theory of Dirac spinors. There occurs a conceptual problem related to obtaining the energy-momentum tensor of the spinor matter field from the variation of the matter field Lagrangian. For the scalar or tensor fields, energy-momentum tensor is the quantity describing the reaction of the matter field Lagrangian to the variations of the metric, while the matter field is held constant during the change of the metric. But for the spinor fields, the above procedure does not hold for obtaining the energy-momentum tensor from the variation with respect to metric only, because the spinor fields are the sections of a spinor bundle obtained as an associated vector bundle from the bundle of spin frames. The bundle of spin frames is a double covering of the bundle of oriented and time-oriented orthonormal frames. For spinor fields, when one varies the metric, the components of the spinor fields cannot be held fixed with respect to some fixed holonomic frame induced by a coordinate system, as in the tensor field case [64]. Therefore, the intrinsic spin of matter field in curved space time requires ECSK theory which is the simplest generalization of the metric-affine formulation of general relativity.

According to the metric-affine formulation of the gravity, the dynamical variables are the tetrad (vierbein, or frame) field $e_a^i$ and the spin connection $\omega_{bk}^a = e_j^a e_{b;k}^j = e_j^a(e_{b,k}^j + \Gamma_{ik}^j e_b^i)$. Here comma denotes the partial derivative with respect to the $x^k$ coordinate, while the semicolon refers to the covariant derivative with respect to the affine connection $\Gamma_{jk}^i$. The antisymmetric lower indices of the affine connection give the torsion tensor $S_{jk}^i = \Gamma_{[jk]}^i$. The tetrad gives the relation between spacetime coordinates denoted by the indices $i, j, k, ...$ and local Lorentz coordinates denoted by the indices $a, b, c, ...$, such that $V^a = V^i e_i^a$, where $V^a$ is a Lorentz vector and $V^i$ is a usual vector. Covariant derivative of a Lorentz vector is defined with respect to the spin connection and denoted by a bar, $V_{|i}^a = V_{,i}^a + \omega_{bi}^a V^b$ and $V_{a|i} = V_{a,i} - \omega_{ai}^b V_b$. Also the covariant derivative of a vector is defined in terms of the affine connection, $V_{;i}^k = V_{,i}^k + \Gamma_{li}^k V^l$ and $V_{k;i} = V_{k,i} - \Gamma_{ki}^l V_l$. Local Lorentz coordinates are lowered or raised by the Minkowski metric $\eta_{ab}$ of the flat spacetime, while the spacetime coordinates are lowered or raised by the metric tensor $g_{ik}$. Metric compatibility condition $g_{ij;k} = 0$ leads the definition of affine connection $\Gamma_{ij}^k = \{_{ij}^k\} + C_{ij}^k$ in terms of the Christoffel symbols $\{_{ij}^k\} = (1/2)g^{km}(g_{mi,j} + g_{mj,i} - g_{ij,m})$ and the contortion tensor $C_{jk}^i = S_{jk}^i + 2S_{(jk)}^i$. Throughout the paper, the $A_{(jk)} = (1/2)(A_{jk} + A_{kj})$ notation is used for symmetrization and $A_{[jk]} = A_{jk} - A_{kj}$ is for the antisymmetrization. With the definitions $g_{ik} = \eta_{ab}e_i^a e_k^b$ and $S_{ik}^j = \omega_{[ik]}^j + e_{[i,k]}^a e_a^j$, the metric tensor and the torsion tensor can be considered as the dynamical variable instead of the tetrad and spin connection.

A tensor density $\mathrm{A}_{kl..}^{ij...}$ is given in terms of the corresponding tensor $A_{kl..}^{ij...}$ as $\mathrm{A}_{kl..}^{ij...} = e A_{kl..}^{ij...}$, where $e = \det e_i^a = \sqrt{-\det g_{ik}}$. Therefore, we represent the spin density and the energy-momentum density, such as $\sigma_{ijk} = e\, s_{ijk}$ and $T_{ik} = e T_{ik}$. Here we call these tensors as metric spin tensor and metric energy-momentum tensor, since the spacetime coordinate indices label these tensors and obtained from the variation of the Lagrangian with respect to the torsion (or contortion) tensor $C_k^{ij}$ and the metric tensor $g^{ij}$, respectively. The metric spin tensor is written as $s_{ij}^k = (2/e)(\delta \ell_m / \delta C_k^{ij}) = (2/e)(\partial \ell_m / \partial C_k^{ij})$, while the metric energy-momentum tensor is given $T_{ij} = (2/e)(\delta \ell_m / \delta g^{ij}) = (2/e)[\partial \ell_m / \partial g^{ij} - \partial_k (\partial \ell_m / \partial (g_{,k}^{ij}))]$. Here, the Lagrangian density of the source matter field is $\ell_m = e L_m$. When the local Lorentz coordinates are also used in these tensors as $\sigma_{ab}^i = e\, s_{ab}^i$ and $T_i^a = e T_i^a$, we call $s_{ab}^i$ and $T_i^a$ as dynamical spin tensor and dynamical energy-momentum tensor, respectively, and they are obtained from the variation of the Lagrangian with respect to the tetrad $e_a^i$ and the spin connection $\omega_i^{ab}$. The dynamical spin tensor is $s_{ab}^i = (2/e)(\delta \ell_m / \delta \omega_i^{ab}) = (2/e)(\partial \ell_m / \partial \omega_i^{ab})$, and energy-momentum tensor is $T_i^a = (1/e)(\delta \ell_m / \delta e_a^i) = (1/e)[\partial \ell_m / \partial e_a^i - \partial_j (\partial \ell_m / \partial (e_{a,j}^i))]$.

Total action of the gravitational field and the source matter in metric-affine ECSK theory is given in the same form with the classical Einstein-Hilbert action, such as $S = \kappa \int (\ell_g + \ell_m) d^4 x$, where $\kappa = 8\pi G$ and $\ell_g = -(1/2\kappa) e R$ is the gravitational Lagrangian density. Here Ricci scalar is constructed from the spin connection containing curvature tensor, such that $R = R_j^b e_b^j$ where $R_j^b = R_{jk}^{bc} e_c^k$ is the Ricci tensor obtained from the curvature tensor $R_{jk}^{bc}$ and finally this curvature tensor is expressed in terms of the spin connection, such that $R_{bij}^a = \omega_{bj,i}^a - \omega_{bi,j}^a + \omega_{ci}^a \omega_{bj}^c - \omega_{cj}^a \omega_{bi}^c$. Variation of the total action with respect to the contortion tensor gives Cartan equations $S_{ik}^j - S_i\, \delta_k^j + S_k\, \delta_i^j = -(\kappa/2e)\sigma_{ik}^j$ and with respect to the metric tensor gives Einstein equations in the form of $G_{ik} = \kappa(T_{ik} + U_{ik})$ where $G_{ik} = P_{ijk}^j - (1/2) P_{lm}^{lm} g_{ik}$ is the Einstein tensor and $P_{ijk}^j$ is the Riemann curvature tensor satisfying the equation $R_{klm}^i = P_{klm}^i + C_{km:l}^i - C_{kl:m}^i + C_{km}^j C_{jl}^i - C_{kl}^j C_{jm}^i$ where colon denotes the Riemannian covariant derivative with respect to the Levi-Civita connection, such as $V_{:i}^k = V_{,i}^k + \{_{li}^k\} V^l$ and $V_{k:i} = V_{k,i} - \{_{ki}^l\} V_l$. Also for torsion-free general relativity theory, curvature tensor turns out to be the Riemann tensor. Right hand side of Eintein equations contains an extra term $U_{ik}$ which is the contribution to the energy-momentum tensor from the torsion and it is quadratic in the spin tensor, such as $U_{ik} = \kappa(-s^{ij}_{[l}s^{kl}_{j]} - (1/2)s^{ijl}s^{k}_{jl} + (1/4)s^{jli}s^{k}_{jl} + (1/8)g^{ik}(-4s^{l}_{j[m}s^{jm}_{l]} + s^{jlm}s_{jlm}))$. Therefore, the total energy-momentum tensor is $\Theta_{ik} = T_{ik} + U_{ik}$.

In metric-affine ECSK formulation of gravity, a spinor quintom field with intrinsic spin has a Lagrangian density of the form $\ell_m = e(i/2)(\bar{\psi}\gamma^k \psi_{;k} - \bar{\psi}_{;k}\gamma^k\psi) - eV$, where $V$ is the potential of the spinor field $\psi$ and the adjoint spinor $\bar{\psi} = \psi^+ \gamma^0$. The covariant derivative of the spinor field is given as $\psi_{;k} = \psi_{,k} - \Gamma_k \psi$ and $\bar{\psi}_{;k} = \bar{\psi}_{,k} - \Gamma_k \bar{\psi}$, where $\Gamma_k = -(1/4)\omega_{abk}\gamma^a\gamma^b$ is the Fock-Ivanenko spin connection, then $\gamma^k$ and $\gamma^a$ are the metric and dynamical Dirac gamma matrices satisfying $\gamma^k = e^k_a \gamma^a$, $\gamma^{(k}\gamma^{m)} = g^{km}I$ and $\gamma^{(a}\gamma^{b)} = \eta^{ab}I$. The covariant derivative of the spinor can be decomposed into the Riemannian covariant derivative plus a contortion tensor $C_{ijk}$ containing term, such as $\psi_{;k} = \psi_{:k} + (1/4)C_{ijk}\gamma^{[i}\gamma^{j]}\psi$ and $\bar{\psi}_{;k} = \bar{\psi}_{:k} - (1/4)C_{ijk}\bar{\psi}\gamma^{[i}\gamma^{j]}$. The Riemannian covariant derivative of the spinor and adjoint spinor fields for quintom DE are given: $\psi_{:k} = \psi_{,k} + (1/4)g_{ik}\{^i_{jm}\}\gamma^j\gamma^m\psi$ and $\bar{\psi}_{:k} = \bar{\psi}_{,k} - (1/4)g_{ik}\{^i_{jm}\}\gamma^j\gamma^m\bar{\psi}$. These covariant derivatives including the contortion tensor $C_{ijk}$ are embedded in the spinor quintom Lagrange density. However, it is needed the explicit form of the contortion tensor which can be obtained from the Cartan equations. Since the right hand side of Cartan equations contains the spin tensor density, we obtain the spin tensor from the variation of the spinor Lagrangian with respect to the contortion tensor, such that $s^{ijk} = (1/e)\sigma^{ijk} = -(1/e)\varepsilon^{ijkl}s_l$, where $\varepsilon^{ijkl}$ is the Levi-Civita symbol, $s^i = (1/2)\bar{\psi}\gamma^i\gamma^5\psi$ is the spin pseudovector, and $\gamma^5 = i\gamma^0\gamma^1\gamma^2\gamma^3$. Inserting the spin tensor for spinor quintom field in the Cartan equations gives the torsion tensor $S_{ijk} = C_{ijk} = (1/2)\kappa\varepsilon_{ijkl}s^l$ which will takes place in the spinor quintom Lagrange density [53-63].

The variation of the spinor quintom matter Lagrangian density with respect to the adjoint spinor gives the ECSK Dirac equation, such as

$$i\gamma^k\psi_{;k} - \frac{\partial V}{\partial \bar{\psi}} + \frac{3}{8}\kappa(\bar{\psi}\gamma^k\gamma^5\psi)\gamma_k\gamma^5\psi = 0, \tag{1}$$

and the variation with respect to the spinor itself gives adjoint Dirac equation as

$$i\overline{\psi}_{:k}\gamma^k + \frac{\partial V}{\partial \psi} - \frac{3}{8}\kappa(\overline{\psi}\gamma^k\gamma^5\psi)\overline{\psi}\gamma_k\gamma^5 = 0. \tag{2}$$

Then the total energy-momentum tensor of the spinor quintom field is obtained from $\Theta_{ik} = T_{ik} + U_{ik}$. Here the metric energy-momentum is obtained by the variation of spinor quintom Lagrange density with respect to the metric tensor, such as

$$T_{ik} = \frac{2}{e}\left[\frac{\partial \ell_m}{\partial g^{ik}} / -\partial_j\left(\frac{\partial \ell_m}{\partial(g^{ik}_{,j})}\right)\right] = \frac{i}{2}(\overline{\psi}\delta^j_{(i}\gamma_{k)}\psi_{;j} - \overline{\psi}_{;j}\delta^j_{(i}\gamma_{k)}\psi) - \frac{i}{2}(\overline{\psi}\gamma^j\psi_{;j} - \overline{\psi}_{;j}\gamma^j\psi)g_{ik} + Vg_{ik}, \tag{3}$$

and the spin contributing metric energy-momentum tensor is obtained by substituting the spin tensor for spinor quintom field in $U_{ik}$. Then the total metric energy-momentum tensor is found to be

$$\Theta_{ik} = \frac{i}{2}(\overline{\psi}\delta^j_{(i}\gamma_{k)}\psi_{:j} - \overline{\psi}_{:j}\delta^j_{(i}\gamma_{k)}\psi) + \frac{3}{4}\kappa s^l s_l g_{ik}. \tag{4}$$

Here, the semicolon covariant derivatives of the spinor field in (3) is decoupled into colon covariant derivatives in (4) and the contortion tensor containing parts of the decoupled covariant derivatives are suppressed in the spin pseudovector $s^l$ by the contribution of $U_{ik}$. In order to rewrite (4) in a more convenient form for our further calculations, we multiply (1) by adjoint spinor $\overline{\psi}$ from the left, and multiply (2) by spinor $\psi$ from right, such that

$$i\overline{\psi}\gamma^k\psi_{:k} - V'\overline{\psi}\psi + \frac{3}{8}\kappa(\overline{\psi}\gamma^k\gamma^5\psi)(\overline{\psi}\gamma_k\gamma^5\psi) = 0, \tag{5}$$

$$i\overline{\psi}_{:k}\gamma^k\psi + V'\overline{\psi}\psi - \frac{3}{8}\kappa(\overline{\psi}\gamma^k\gamma^5\psi)(\overline{\psi}\gamma_k\gamma^5\psi) = 0, \tag{6}$$

where $V' = \partial V/\partial(\overline{\psi}\psi)$ for which $\overline{\psi}(\partial V/\partial\overline{\psi}) = (\partial V/\partial\psi)\psi = V'\overline{\psi}\psi$. By using (5) and writing the symmetrizations explicitly in (4), we obtain the total energy-momentum tensor $\Theta_{ik}$ of the spinor field dark energy in the form of

$$\Theta_{ik} = \frac{i}{4}(\overline{\psi}\gamma_i \psi_{:k} + \overline{\psi}\gamma_k \psi_{:i} - \overline{\psi}_{:i}\gamma_k \psi - \overline{\psi}_{:k}\gamma_i \psi) + \frac{1}{2}(V'\overline{\psi}\psi - i\overline{\psi}\gamma^l \psi_{:l})g_{ik}. \tag{7}$$

We consider the spinor quintom DE model in a FRW spacetime whose metric is given as

$$ds^2 = dt^2 - a^2(t)[dx^2 + dy^2 + dz^2], \tag{8}$$

and the corresponding tetrad components read

$$e_0^i = \delta_0^i, \qquad e_a^i = \frac{1}{a(t)}\delta_a^i. \tag{9}$$

Therefore, by performing the Riemannian covariant derivatives explicitly in (7), the timelike components $\Theta_{00}^\psi$ and the spacelike components $\Theta_{u}^\psi$ of the space independent spinor field dark energy energy-momentum tensor can be obtained, such as

$$\Theta_{00} = -\frac{i}{2}\dot{\overline{\psi}}\gamma_0\psi + \frac{1}{2}V'\overline{\psi}\psi - \frac{3i}{8}H\overline{\psi}\gamma_0\psi, \tag{10}$$

$$\Theta_{u} = -\frac{i}{2}\overline{\psi}\gamma_0\dot{\psi}\, g_u + \frac{1}{2}V'\overline{\psi}\psi\, g_u - \frac{3i}{8}H\overline{\psi}\gamma_0\psi\, g_u, \tag{11}$$

Here $H = \dot{a}(t)/a(t)$ is the Hubble parameter and it comes from the Levi-Civita connections in the Riemannian covariant derivatives. We now write the ECSK Dirac equation (4) and (5) for a space independent spinor field as

$$i\overline{\psi}\gamma^0\dot{\psi} + \frac{3i}{4}H\overline{\psi}\gamma^0\psi - V'\overline{\psi}\psi + \frac{3}{8}\kappa(\overline{\psi}\gamma^0\gamma^5\psi)(\overline{\psi}\gamma_0\gamma^5\psi) = 0, \tag{12}$$

$$i\dot{\overline{\psi}}\gamma^0\psi + \frac{3i}{4}H\overline{\psi}\gamma^0\psi + V'\overline{\psi}\psi - \frac{3}{8}\kappa(\overline{\psi}\gamma^0\gamma^5\psi)(\overline{\psi}\gamma_0\gamma^5\psi) = 0. \tag{13}$$

The solution of (12) and (13) by adding them leads

$$\overline{\psi}\dot{\psi} + \dot{\overline{\psi}}\psi + \frac{3}{2}H\overline{\psi}\psi = 0, \tag{14}$$

and

$$\overline{\psi}\psi = \frac{N}{a^{3/2}}. \tag{15}$$

Here $N$ is the integration constant, then by using the scale factor $a \propto e^{\alpha t}$ for a cosmological fluid [65], we can also obtain $\overline{\psi}\psi = N e^{-3\alpha t/2}$. Using (13) in (10) leads to the energy density

$$\rho = \Theta_0^0 = V'\overline{\psi}\psi - \frac{3}{16}\kappa(\overline{\psi}\gamma^0\gamma^5\psi)(\overline{\psi}\gamma_0\gamma^5\psi), \tag{16}$$

and similarly using (12) in (11) leads to the pressure of the spinor field dark energy

$$p = -\Theta_t^t = -\frac{3}{16}\kappa(\overline{\psi}\gamma^0\gamma^5\psi)(\overline{\psi}\gamma_0\gamma^5\psi), \tag{17}$$

respectively. Then the EoS of the spinor field is given as

$$\omega = \frac{p}{\rho} = \frac{(3/16)\kappa(\overline{\psi}\gamma^0\gamma^5\psi)^2}{(3/16)\kappa(\overline{\psi}\gamma^0\gamma^5\psi)^2 - V'\overline{\psi}\psi}, \tag{18}$$

where $\gamma^0 = \gamma_0$ for a FRW metric. We rewrite the EoS in the form of

$$\omega = -1 + \alpha, \tag{19}$$

where

$$\alpha = \frac{6\kappa(\overline{\psi}\gamma^0\gamma^5\psi)^2 - 16V'\overline{\psi}\psi}{3\kappa(\overline{\psi}\gamma^0\gamma^5\psi)^2 - 16V'\overline{\psi}\psi}, \tag{20}$$

It is known that for $\alpha = 4/3$ the EoS of the spinor field is $\omega = 1/3$ and it behaves like radiation, but for $\alpha = 1$, $\omega = 0$ and it is normal matter. On the other hand, if $\alpha < 2/3$ the EoS $\omega < -1/3$ meaning that the spinor field behaves like a DE leading to the acceleration of universe. The $\alpha < 2/3$ region allows us to investigate the dynamical evolution of the spinor quintom DE described in ECSK formalism with intrinsic spin.

## 3. Dynamical evolution of spinor quintom

From (20) we deduce that the spinor field can have an EoS of $-1<\omega<-1/3$ for $0<\alpha<2/3$ and shows a quintessence-like behavior, but it has $\omega=-1$ Cosmological constant value if $\alpha=0$, then it behaves like a phantom for $\omega<-1$ if $\alpha<0$. Therefore, the spinor field exhibits a quintom picture by crossing the Cosmological constant boundary $\omega=-1$ from above, or below this boundary depending on the sign of $\alpha$ in (20).

There exist three categories of spinor quintom evolution depending on the behavior of the potential $V$. The quintom scenario may exhibit an evolution starting from $-1<\omega$ quintessence phase to $\omega<-1$ phantom phase, called as Quintom-A. Other scenario may evolve from $\omega<-1$ to $-1<\omega$, Quintom-B scenario. The last quintom scenario contains the evolution in which crossing $\omega=-1$ more than one time, called as Quintom-C model.

Considering the quintom scenario in which the spinor field comes from $-1<\omega$ quintessence phase to $\omega<-1$ phantom phase, we first need to find the condition $0<\alpha$. Since the energy density (16) must be positive definite, $V'$ is positive. Therefore, the condition of occurring $-1<\omega$ phase reads from (20) as

$$16V'\bar{\psi}\psi > 6\kappa(\bar{\psi}\gamma^0\gamma^5\psi)^2. \qquad (21)$$

Similarly, $\omega=-1$ boundary occurs for

$$16V'\bar{\psi}\psi = 6\kappa(\bar{\psi}\gamma^0\gamma^5\psi)^2, \qquad (22)$$

and $\omega<-1$ phantom phase occurs for

$$16V'\bar{\psi}\psi < 6\kappa(\bar{\psi}\gamma^0\gamma^5\psi)^2, \qquad (23)$$

Since prime denotes the derivative with respect to $\bar{\psi}\psi$, the solution of (22) is found as $V_\Lambda=(6\kappa/16)(\bar{\psi}\gamma^0\gamma^5\psi)^2\ln\bar{\psi}\psi$, which the dynamical evolution of potential goes to the Cosmological constant boundary.

In order to obtain a Quintom-A scenario, we define the potential to be

$$V = (6\kappa/16)(\bar{\psi}\gamma^0\gamma^5\psi)^2 \ln \bar{\psi}\psi - (c - \bar{\psi}\psi)\bar{\psi}\psi, \qquad (24)$$

for the early times of the universe. Then the potential leads the EoS from (20) as

$$\omega = -1 + \frac{32(c - \bar{\psi}\psi)\bar{\psi}\psi}{32(c - \bar{\psi}\psi)\bar{\psi}\psi + (3\kappa/16)(\bar{\psi}\gamma^0\gamma^5\psi)^2}, \qquad (25)$$

the term $(2\bar{\psi}\psi - c)$ in this potential satisfies $-1 < \omega$ quintessence scenario (21) with (15), since the scaling factor $a$ is very small at the beginning of the evolution of the universe. When $\bar{\psi}\psi$ becomes equal to $c/2$, this potential leads the spinor field to approach $\omega = -1$ boundary (22). After that scaling factor evolves to a greater value, then $\bar{\psi}\psi$ reaches a value smaller than $c/2$. This gives the condition (23) phantom phase $\omega < -1$. We illustrate this behavior in Figure 1 by numerical analysis. According to the figure, $\omega$ starts its evolution from above $-1$ to below $-1$. We set the crossing Cosmological constant boundary as at $t = 0$. After crossing the $-1$ boundary, spinor quintom picks up and avoid from a Big Rip singularity then enters a stable matter dominated expansion with $\omega = 0$ value.

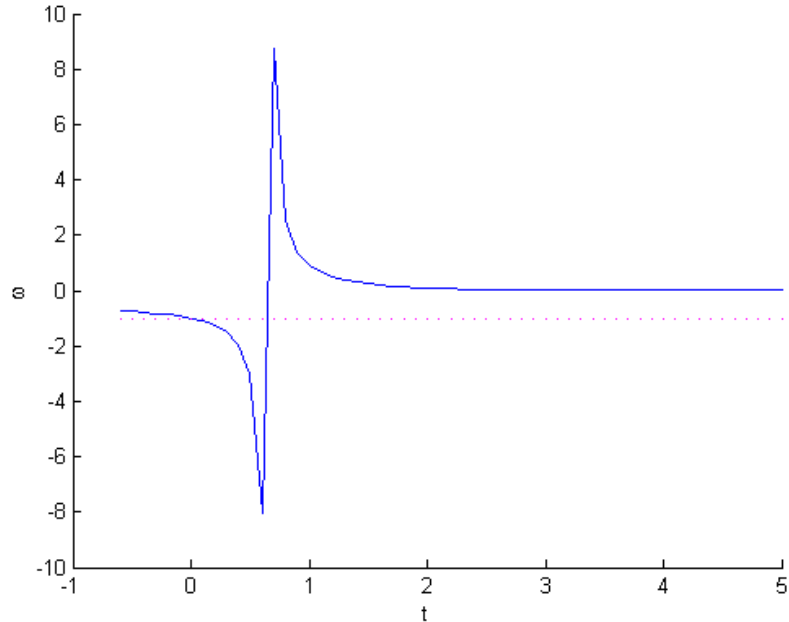

Figure 1. Evolution of $\omega$ with potential (24) as a function of time. For the numerical analysis we assume $N = 3$, $c = 6$, $\alpha = 1$. From Ref. [52].

For a Quintom-B model, the potential can be defined as

$$V = (6\kappa/16)(\bar{\psi}\gamma^0\gamma^5\psi)^2 \ln\bar{\psi}\psi - (c-\bar{\psi}\psi)^2, \tag{26}$$

then the EoS is obtained, such that

$$\omega = -1 + \frac{16(2\bar{\psi}\psi - c)\bar{\psi}\psi}{16(2\bar{\psi}\psi - c)\bar{\psi}\psi + (3\kappa/16)(\bar{\psi}\gamma^0\gamma^5\psi)^2}, \tag{27}$$

which lead to $\omega < -1$ phantom phase (23) at the beginning of the evolution of universe. With the increasing of the scale factor, $\bar{\psi}\psi$ decreases to $c$ and the term $2(c-\bar{\psi}\psi)$ becomes zero. This gives $\omega = -1$ Cosmological constant phase (22). As the evolution continues $\bar{\psi}\psi$ gets smaller than $c$ and spinor quintom reaches a quintessence scenario $-1 < \omega$ in (21). The behavior of the spinor Quintom-B scenario is represented in Figure 2 which states that the spinor field starts the evolution from below $\omega = -1$ to above $\omega = -1$. Crossing from phantom to quintessence phase continues in this phase with an EoS value of $-1 < \omega < -1/3$ which imitates a stable de Sitter accelerated expansion for a scalar field dark energy model.

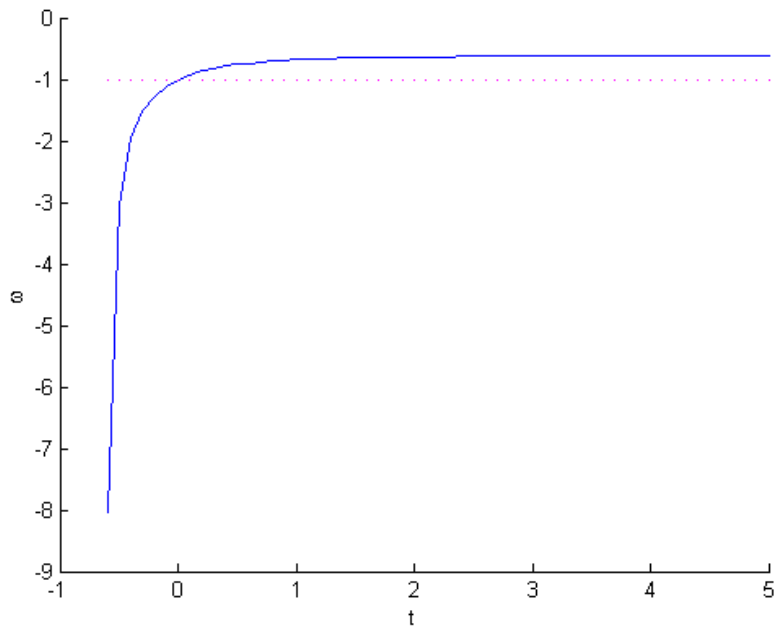

Figure 2. Evolution of $\omega$ with potential (26) as a function of time. For the numerical analysis we assume $N = 3$, $c = 3$, $\alpha = 1$. From Ref. [52].

Third case Quintom-C scenario can be obtained for the potential

$$V = (6\kappa/16)(\bar{\psi}\gamma^0\gamma^5\psi)^2 \ln \bar{\psi}\psi - (c - \bar{\psi}\psi)^2 \bar{\psi}\psi, \tag{28}$$

which leads the EoS as

$$\omega = -1 + \frac{16(c - 3\bar{\psi}\psi)(\bar{\psi}\psi - c)\bar{\psi}\psi}{16(c - 3\bar{\psi}\psi)(\bar{\psi}\psi - c)\bar{\psi}\psi + (3\kappa/16)(\bar{\psi}\gamma^0\gamma^5\psi)^2}. \tag{29}$$

This potential provides two roots in $V'\bar{\psi}\psi$ for crossing the $\omega = -1$ boundary. The term coming from the derivative of $V$ is $-3(\bar{\psi}\psi)^2 + 4c\bar{\psi}\psi - c^2$ which determines the sign of $16V'\bar{\psi}\psi$ in (21)-(23). During the evolution of universe with the increase in scale factor, $\bar{\psi}\psi$ decreases firstly to the value $c$ which is the bigger root. This is a transition from phantom phase to quintessence phase by crossing $-1$ boundary. After continuing the evolution $\bar{\psi}\psi$ decreases to the second root $c/3$ which is re-crossing the $-1$ boundary as a transition from quintessence phase to phantom phase again. This scenario is obviously a Quinton-C scenario and is illustrated in Figure 3. We see from the figure that the EoS of the quintom model crosses the $\omega = -1$ boundary twice, firstly from below $\omega = -1$ to above $\omega = -1$, secondly from above to below $\omega = -1$, then it picks up then avoid from Big Rip singularities and finally it asymptotically evolves to a stable matter dominated expansion epoch with a value of $\omega = 0$.

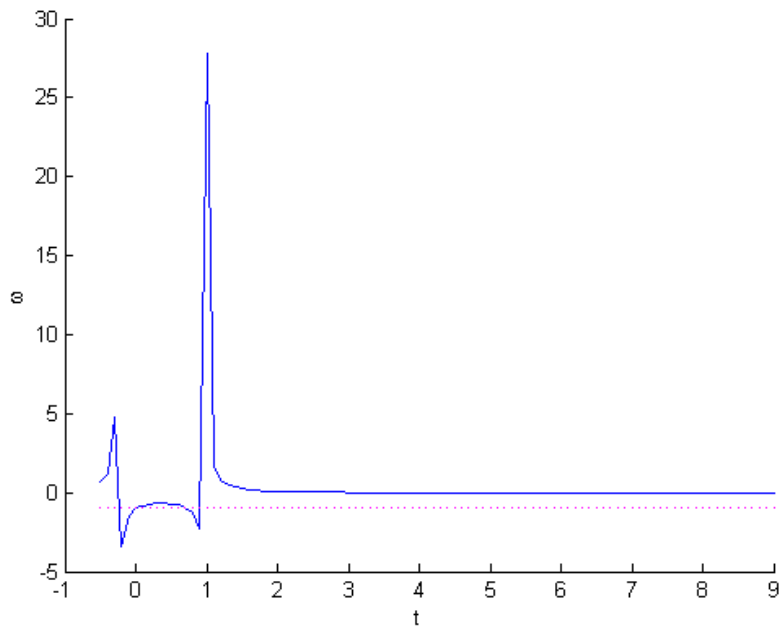

Figure 3. Evolution of $\omega$ with potential (24) as a function of time. For the numerical analysis we assume $N = 3$, $c = 3$, $\alpha = 1$. From Ref. [52].

Although considering the phantom scenarios normally leads to the Big Rip singularities due to the unbound of EoS from below $\omega=-1$, our spinor quintom model with intrinsic spin in ECSK theory avoids from the Big Rip singularities by picking up to a bound value and approaching to a stable value, as seen in Figure 1 and 3. Diverging EoS of a dark fluid from a constant bound toward a lower singularity refers to continuous increase in the pressure of the fluid. This scenario is avoided in spinor quintom with intrinsic spin, which may be interpreted as the intrinsic spin of the fluid quanta leads to a bound pressure value. The increase of the pressure with the energy density is bounded due to the effect of intrinsic spin, then singular values of EoS are avoided, and the universe enters a stable expansion in the final era.

## 4. Conclusion

By using the spinor field dark energy in a FRW geometry, a consistent quintom model in which EoS crosses $-1$ boundary without using a ghost field, has recently been obtained in the framework of general relativity [52]. Here, we consider the spinor field dark energy with intrinsic spin in the formalism of metric-affine ECSK theory. We first introduce the ECSK formalism, then define the model Lagrangian whose variations with respect to the tetrad field and torsion tensor gives the total energy-momentum tensor consisting of metric and spin contributions. Also from the variation of Lagrangian with respect to the spinor field we obtain the ECSK Dirac equation. By using the total energy momentum tensor and ECSK Dirac equation, the energy density and the pressure values of the spinor quintom DE is obtained, from which the EoS of the model is obtained for an arbitrary potential. The dependence of the potential on the spinor field leads to the evolution of potential with the change of scale factor, since the scale factor is increases by time. Constructing the ECSK spinor potential suitably the quintom scenario is reached, for three different cases as Quintom A, B and C models.

The Quintom-A case exhibits the transition of EoS from quintessence phase to phantom phase, evolving to a stable matter dominated expansion with $\omega \to 0$. This scenario is avoided from the Big Rip singularities due to the balancing of energy density and pressure of spinor DE by intrinsic spin. Similarly, in the Quintom-B scenario, the EoS of the model evolves from phantom region $\omega<-1$ to quintessence region $\omega>-1$, and approaches an EoS value of $-1<\omega<-1/3$ referring to a stable de Sitter accelerated expansion for a scalar field dark energy model. On the other hand, the Quintom-C scenario exhibits the evolution of EoS which crossed the Cosmological constant boundary $\omega=-1$ more than one time. The spinor

Quintom-C firstly crosses the $-1$ boundary from phantom epoch, and then it again enters the phantom epoch from quintessence epoch. Then it converges to $\omega=0$ stable matter dominated expansion phase by picking up from avoiding the singularities.

The proposed ECSK spinor quintom model differs from the spinor quintom model in the framework of general relativity with the existence of matter dominated expansion phases in cases A and C. In both Quintom-A and Quintom-C cases, after the spinor field crosses the $-1$ boundary from a quintessence epoch toward the phantom epoch, it suddenly picks and enters the stable matter dominated expansion with $\omega=0$. This can be interpreted as the intrinsic spin causes to fix the pressure of the fluid to a certain value as the energy density increases. After the spinor field reaches a very large energy density value, this allows neglecting the pressure relative to energy density value, which imitates a pressure free matter dominated era with zero EoS.

**References**

[1] S. Perlmutter, G. Aldering, G. Goldhaber et al., *The Astrophysical Journal*, 517, 2, 565, (1999).

[2] A. G. Riess, A. V. Filippenko, P. Challis et al., *The Astronomical Journal*, 116, 3, 1009, (1998).

[3] U. Seljak, A. Makarov, P. McDonald et al., *Phys. Rev. D*, 71, 103515, (2005).

[4] M. Tegmark, M. Strauss, M. Blanton et al., *Phys. Rev. D*, 69, 103501, (2004).

[5] D. J. Eisenstein, I. Zehavi, D. W. Hogg et al., *The Astrophysical Journal*, 633, 2, 560, (2005).

[6] D. N. Spergel, L. Verde, H. V. Peiris et al., *The Astrophysical Journal Supplement Series*, 148, 1, 175, (2003).

[7] E. Komatsu, K. M. Smith, J. Dunkley et al., *The Astrophysical Journal Supplement Series*, 192, 2, (2011).

[8] G.Hinshaw, D. Larson, E. Komatsu et al., *The Astrophysical Journal Supplement Series*, 208, 2, 19, (2013).

[9] P.A.R. Ade, N. Aghanim, C. Armitage-Caplan, *Astronomy & Astrophysics*, (2013).

[10] S. Weinberg, *Rev. Modern Phys.*, 61, 1, (1989).

[11] S.M. Carroll, W.H. Press, E.L. Turner, *Ann. Rev. Astron. Astrophys.*, 30, 499, (1992).

[12] L.M. Krauss, M.S. Turner, *Gen. Relativity Gravitation,* 27, 1137, (1995).

[13] G. Huey, L.M. Wang, R. Dave, R.R. Caldwell, P.J. Steinhardt, *Phys. Rev. D*, 59 063005, (1999).

[14] B. Ratra, P.J.E. Peebles, Phys. Rev. D 37, 3406, (1988).

[15] C. Wetterich, Nuclear Phys. B 302, 668, (1988).

[16] R.R. Caldwell, M. Kamionkowski, N.N. Weinberg, Phys. Rev. Lett. 91, 071301, (2003).

[17] R.R. Caldwell, Phys. Lett. B 545, 23 (2002).

[18] S.M. Carroll, M. Hoffman, M. Trodden, Phys. Rev. D 68, 023509, (2003).

[19] R.R. Caldwell, M. Kamionkowski, N.N. Weinberg, Phys. Rev. Lett. 91, 071301, (2003).

[20] Y.F. Cai, , E.N. Saridakis, M.R. Setare, J.Q. Xia, Physics Reports, 493, 1-60, (2010).

[21] D. Huterer, A. Cooray, Phys. Rev. D 71, 023506, (2005).

[22] B. Feng, X.L. Wang, X.M. Zhang, Phys. Lett. B 607, 35, (2005).

[23] E. Komatsu, et al., Astrophys. J. Suppl. 180, 330, (2009).

[24] J.Q. Xia, H. Li, G.B. Zhao, X. Zhang, Phys. Rev. D 78, 083524, (2008).

[25] H. Li, J. Liua, J.Q. Xiac, L. Sund, Phys. Lett. B, 675, 2, (2009).

[26] J.Q. Xia, Y.F. Cai, T.T. Qiu, G.B. Zhao, J. Modern Phys. D 17, 1229, (2008).

[27] A. Vikman, Phys. Rev. D 71, 023515, (2005).

[28] W. Hu, Phys. Rev. D 71, 047301, (2005).

[29] R.R. Caldwell, M. Doran, Phys. Rev. D 72, 043527, (2005).

[30] G.B. Zhao, J.Q. Xia, M. Li, B. Feng, X. Zhang, Phys. Rev. D 72, 123515, (2005).

[31] M. Kunz, D. Sapone, Phys. Rev. D 74, 123503, (2006).

[32] B. Feng, X.-L. Wang and X.-M. Zhang, Phys. Lett. B 607, 35, (2005).

[33] S. Nojiri, S.D. Odintsov and M. Sami, Phys. Rev. D 74, 046004, (2006).

[34] A. Vikman, Phys. Rev. D 71, 023515, (2005).

[35] M.-z. Li, B. Feng and X.-m. Zhang, JCAP 12, 002, (2005).

[36] L. Perivolaropoulos, JCAP 10, 001, (2005).

[37] J. Sadeghi, M. Setare, A. Banijamali and F. Milani, Phys. Rev. D 79, 123003, (2009).

[38] V.K. Onemli, R.P. Woodard, Classical Quantum Gravity 19, 4607, (2002).

[39] V.K. Onemli, R.P. Woodard, Phys. Rev. D 70, 107301, (2004).

[40] T. Brunier, V.K. Onemli, R.P. Woodard, Classical Quantum Gravity 22, 59, (2005).

[41] E.O. Kahya, V.K. Onemli, Phys. Rev. D 76, 043512, (2007).

[42] E.O. Kahya, V.K. Onemli, R.P. Woodard, Phys. Rev. D 81, 023508, (2010).

[43] A.H. Taub, Phys. Rev. 51, 512, (1937).

[44] D.R. Brill, J.A. Wheeler, Rev. Modern Phys. 29, 465, (1957).

[45] L. Parker, Phys. Rev. D 3, 346, (1971); 3, 2546, (1971).

[46] Y.N. Obukhov, Phys. Lett. A 182, 214, (1993).

[47] C. Armendariz-Picon, P.B. Greene, Gen. Relativity Gravitation 35, 1637, (2003).

[48] S. Kasuya, Phys. Lett. B 515, 121, (2001).

[49] B. Saha, Modern Phys. Lett. A 16, 1287, (2001).

[50] M.O. Ribas, F.P. Devecchi, G.M. Kremer, Phys. Rev. D 72, 123502, (2005).

[51] L.P. Chimento, F.P. Devecchi, M. Forte, G.M. Kremer, Classical Quantum Gravity 25, 085007, (2008).

[52] Y.F. Cai, J. Wang, Classical Quantum Gravity 25, 165014, (2008).

[53] E. A. Lord, Tensors, Relativity and Cosmology, (McGraw-Hill, 1976);

[54] T. W. B. Kibble, J. Math. Phys. 2, 212, (1961).

[55] D. W. Sciama, in Recent Developments in General Relativity, (Pergamon, 1962)

[56] V. de Sabbata and C. Sivaram, Spin and Torsion in Gravitation, (World Scientific, 1994).

[57] R. Utiyama, Phys. Rev. 101, 1597, (1956).

[58] T. P. Sotiriou, V. Faraoni, Rev. Mod. Phys. 82, 451, (2010).

[59] S. Capozziello, M. De Laurentis, Phys. Rept. 509, 167, (2011).

[60] N. J. Poplawski, Phys. Lett. B 690, 73 (2010).

[61] I. L. Shapiro, Phys. Rep. 357, 113, (2002).

[62] F. W. Hehl, B. K. Datta, J. Math. Phys. 12, 1334, (1971).

[63] N. Poplawski, Intrinsic spin requires gravity with torsion and curvature, arXiv:1304.0047 [gr-qc]

[64] M. Forger, H. Römer, Annals of Physics, 309, 2, (2004).

[65] S. Carroll, Spacetime and geometry: An introduction to general relativity, (Addison Wesley 2004).